\documentstyle[prd,aps]{revtex}
\tighten
\draft
\begin{document} 
\bibliographystyle{prsty}

 \mathchardef\ddash="705C

\title{
New $\ddash$Electric" Description of\\
Supersymmetric Quantum Chromodynamics
\footnote{\it To be published in Phys. Lett. B.}
}

\author{Yasuharu Honda$^{a}$
\footnote{E-mail:8jspd005@keyaki.cc.u-tokai.ac.jp}
and Masaki Yasu${\grave {\rm e}}^{a,b}$
\footnote{E-mail:yasue@keyaki.cc.u-tokai.ac.jp}
}

\address{\vspace{5mm}$^{a}${\sl Department of Physics, Tokai University}\\
{\sl 1117 KitaKaname, Hiratsuka, Kanagawa 259-1292, Japan}}
\address{\vspace{2mm}$^{b}${\sl Department of Natural Science\\School of Marine
Science and Technology, Tokai University}\\
{\sl 3-20-1 Orido, Shimizu, Shizuoka 424-8610, Japan}}
\date{TOKAI-HEP/TH-9903, September, 1999}
\maketitle

\begin{abstract}
Responding to the recent claim that the origin of moduli space may be unstable 
in $\ddash$magnetic" supersymmetric quantum chromodynamics (SQCD) with 
$N_f$ $\leq$ $3N_c/2$ ($N_c>2$) for $N_f$ flavors and $N_c$ colors of quarks, 
we explore the possibility of finding nonperturbative physics for $\ddash$electric" 
SQCD.  We present a recently discussed effective superpotential 
for $\ddash$electric" SQCD with $N_c+2$ $\leq$ $N_f$ $\leq$ $3N_c/2$ ($N_c>2$) 
that generates chiral symmetry breaking with a residual nonabelian symmetry 
of $SU(N_c)_{L+R}$ $\times$ $SU(N_f-N_c)_L$ $\times$ $SU(N_f-N_c)_R$.  
Holomorphic decoupling property is shown 
to be respected.  For massive $N_f-N_c$ quarks, our superpotential 
with instanton effect taken into account produces a consistent vacuum structure 
for SQCD with $N_f$ = $N_c$ compatible with the holomorphic decoupling. 
\end{abstract}


It has been widely accepted that physics of $N$=1 supersymmetric quantum 
chromodynamics (SQCD) at strong $\ddash$electric" coupling is well 
described by the corresponding dynamics of SQCD at weak $\ddash$magnetic" 
coupling\cite{Seiberg,EearlySeiberg}.  This dynamical feature is 
referred to as Seiberg's $N$=1 duality\cite{Review}.  In order to apply the $N$=1 
duality to the physics of SQCD, one has to adjust dynamics of $\ddash$magnetic" 
quarks so that the anomaly-matching conditions\cite{tHooft,AnomalyMatch} 
are satisfied. In SQCD with  quarks carrying $N_f$ flavors and $N_c$ colors for $N_f$ 
$\geq$ $N_c$+2, the $N$=1 duality is respected as long as $\ddash$magnetic" 
quarks have $N_f-N_c$ colors\cite{Seiberg,Review}.  Appropriate interactions of 
$\ddash$magnetic" quarks can be derived in SQCD embedded in a softly broken 
$N$=2 SQCD\cite{CoulombPhase,BrokenN_2} that possesses the manifest $N$=2 
duality\cite{SeibergWitten}.  In SQCD with 3$N_c$/2 $<$ $N_f$ $<$ 3$N_c$, its phase 
is characterized by an interacting Coulomb phase\cite{Seiberg}, where the $N$=2 
duality can be transmitted to $N$=1 SQCD. On the other hand, in SQCD with $N_f$ 
$\leq$ 3$N_c$/2, it is not clear that the $N$=1 duality is supported by the similar 
description in terms of the $N$=2 duality although it is believed that the result 
for 3$N_c$/2 $<$ $N_f$ can be safely extended to apply to this case.  

Lately, several arguments\cite{Instability,S,GapEquation} have been made to discuss 
other possibilities than the physics based on the $N$=1 duality, especially for 
SQCD with $N_f$ $\leq$ 3$N_c$/2. It is clamed in Ref.\cite{Instability} that 
the origin of moduli space becomes unstable in 
$\ddash$magnetic" SQCD and that the spontaneous breakdown of the vectorial 
$SU(N_f)_{L+R}$ symmetry is expected to occur.  An idea of an anomalous $U(1)$ 
symmetry, $U(1)_{anom}$, taken as a background gauge symmetry has been employed.  
Their findings are essentially based on the analyses made in the slightly broken 
supersymmetric (SUSY) vacuum.  
On the other hand, emphasizing nonperturbative implementation of  $U(1)_{anom}$, 
the authors of Ref.\cite{S} have derived a new 
type of an effective superpotential applicable to $\ddash$electric" SQCD.  However, 
the physical consequences based on their superpotential 
have not been clarified yet.  Finally, extensive evaluation of formation of condensates has 
provided a signal due to spontaneous breaking of chiral symmetries although there is 
a question on the reliability of their dynamical gap equations\cite{GapEquation}.  
These attempts suggest that, in order to make underlying property of SQCD more transparent,  
it is helpful to employ a composite chiral superfield composed of chiral gauge 
superfields\cite{VY,UsualNfNcConstraint,RecentVY} that is responsible for relevant 
expression for $U(1)_{anom}$. 

In a recent article\cite{Yasue}, we have discussed what physics is suggested by SQCD 
with $N_c+2$ $\leq$ $N_f$ $\leq$ $3N_c/2$ and have found that, once SQCD triggers 
the formation of one condensate made of a quark-antiquark pair, the successive formation of other 
condensates is dynamically induced to generate spontaneous breakdown of the chiral $SU(N_f)$ 
symmetry to $SU(N_f-N_c)$ as a residual chiral nonabelian symmetry.
\footnote{
Another case with a chiral $SU(N_f-N_c+1)$ symmetry has also been discussed.
}
The anomalies associated with original chiral symmetries are matched with those from the 
Nambu-Goldstone superfields.  As in Ref.\cite{Instability}, our suggested dynamics can 
also be made more visible by taking softly broken SQCD\cite{SoftSUSY} in 
its supersymmetric limit.  The derived effective superpotential has the common 
structure to the one discussed in Ref.\cite{S}.   It should be noted that the $\ddash$magnetic" 
description should be selected by SQCD if SQCD favors the formation of no condensates.

In this paper, we further study effects of SUSY-preserving masses. It is shown that our 
superpotential is equipped with holomorphic decoupling property. 
In the case that quarks carrying flavors of $SU(N_f-N_c)$ are massive,  our superpotential 
supplemented by instanton contributions correctly reproduces consistent vacuum structure 
with the decoupling property.  

In SQCD with $N_c+2$ $\leq$ $N_f$ $\leq$ $3N_c/2$ ($N_c$ $>$ 2), 
our superpotential\cite{Yasue} takes the form of
\begin{equation}\label{Eq:Weff}
W_{\rm eff}=S 
\left\{ 
\ln\left[
\frac{
	S^{N_c-N_f}{\rm det}\left(T\right) f(Z)
}
{
	\Lambda^{3N_c-N_f}
} 
\right] 
+N_f-N_c\right\}
\end{equation}
with an arbitrary function, $f(Z)$,  to be determined, 
where $\Lambda$ is the scale of SQCD.  The composite superfields are 
specified by $S$ and $T$:
\begin{equation}\label{Eq:FiledContentST}
S = \frac{1}{32\pi^2}\sum_{A,B=1}^{N_c} W_A^BW_B^A, \
 T_j^i  =  \sum_{A=1}^{N_c} Q_A^i{\bar Q}^A_j,  
\end{equation}
where chiral superfields of quarks and antiquarks are denoted by $Q_A^i$ and ${\bar Q}_i^A$ and 
gluons are by $W_A^B$ with Tr($W$) = 0  for $i$ = 1 $\sim$ $N_f$ and $A$, $B$ = 1 $\sim$ $N_c$. 
The remaining field, $Z$, describes an effective field.   Its explicit form can 
be given by
\begin{equation}\label{Eq:FiledContentZB}
Z = \frac{\sum_{i_1\cdots i_{N_f}, j_1\cdots j_{N_f}} B^{[i_1i_2\cdots i_{N_c}]}T_{j_{N_c+1}}^{i_{N_c+1}}\cdots T_{j_{N_f}}^{i_{N_f}}{\bar B}_{[j_1j_2\cdots j_{N_c}]}}
{{\rm det}\left(T\right)} \bigg( \equiv\frac{BT^{N_f-N_c}{\bar B}}{{\rm det}\left(T\right)}\bigg),
\end{equation}
where
\begin{eqnarray}\label{Eq:FiledContent}
   &  &    B^{[i_1i_2{\dots}i_{N_c}]} = \sum_{A_1\dots A_{N_c}}\frac{1}{N_c!}
                                                 \varepsilon^{A_1A_2{\dots}A_{N_c}}Q_{A_1}^{i_1}\dots 
Q_{A_{N_c}}^{i_{N_c}}, 
 \nonumber \\
    &  &   {\bar B}_{[i_1i_2{\dots}i_{N_c}]} = \sum_{A_1\dots A_{N_c}}\frac{1}{N_c!}
                                                 \varepsilon_{A_1A_2{\dots}A_{N_c}}{\bar Q}_{i_1}^{A_1}\dots 
{\bar Q}_{i_{N_c}}^{A_{N_c}}.
\end{eqnarray}
This superpotential is derived by requiring that 
not only it is invariant under $SU(N_f)_L$ $\times$ $SU(N_f)_R$ as well as under two 
additional $U(1)$ symmetries but also it is equipped with the transformation property 
under $U(1)_{anom}$ broken by the instanton effect, namely, 
$\delta{\cal L}$ $\sim$ $F^{\mu\nu}{\tilde F}_{\mu\nu}$, where ${\cal L}$ 
represents the lagrangian of SQCD and $F^{\mu\nu}$
 (${\tilde F}_{\mu\nu}\sim \epsilon_{\mu\nu\rho\sigma}F^{\rho\sigma}$) 
is a gluon's field strength\cite{VY}. 
Note that $Z$ is neutral under the entire chiral symmetries including $U(1)_{anom}$ and the 
$Z$-dependence of $f(Z)$ cannot be determined by the symmetry principle.  

Although the origin of the moduli space, where $T$ = $B$ = ${\bar B}$ = 0, 
is allowed by $W_{\rm eff}$, the consistent SQCD must automatically show the 
anomaly-matching property with respect to unbroken chiral symmetries. Since 
the anomaly-matching property is not possessed by SQCD realized at $T$ = $B$ = 
${\bar B}$ = 0, composite fields are expected to be dynamically reshuffled so that 
the anomaly-matching becomes a dynamical consequence.   Usually, one accepts that 
SQCD is described by $\ddash$magnetic" degrees of freedom instead of $T$, $B$ 
and ${\bar B}$.  However, it is equally possible to occur that $\ddash$electric" SQCD dynamically 
rearranges some of $T$, $B$ and ${\bar B}$ to develop vacuum expectation values (VEV's).  
In this case, chiral symmetries are spontaneously broken and the presence of the 
anomalies can be ascribed to 
the Nambu-Goldstone bosons for the broken sector and to chiral fermions for 
the unbroken sector\cite{AnomalyMatch}.  If the consistent SQCD with broken chiral 
symmetries is described by our superpotential, the anomaly-matching constraint 
must be automatically satisfied and is indeed shown to be satisfied by the Nambu-Goldstone 
superfields.  In the ordinary QCD with two flavors, we know the similar situation, 
where QCD with massless proton and neutron theoretically allows the existence of 
the unbroken chiral $SU(2)_L$ $\times$ $SU(2)_R$ symmetry but real physics of QCD chooses its 
spontaneous breakdown into $SU(2)_{L+R}$\cite{tHooft}.  However, since the SUSY-invariant theory 
possesses any SUSY vacua, which cannot be dynamically selected, both $\ddash$magnetic" and 
$\ddash$electric" vacua will correspond to a true vacuum of SQCD.  

In the classical limit, where the SQCD gauge coupling $g$ vanishes, the behavior of $W_{\rm eff}$  is readily 
found by applying the rescaling $S$ $\rightarrow$ $g^2S$ and invoking the definition $\Lambda$ $\sim$ 
$\mu\exp (-8\pi^2/(3N_c-N_f)g^2)$, where $\mu$ is a certain reference mass scale. The resulting $W_{\rm eff}$ 
turns out to be $WW/4$, which is the tree superpotential for the gauge kinetic term.  If $S$ is integrated out\cite{Sout}, 
one reaches the ADS-type superpotential\cite{EearlySeiberg}:
\begin{equation}\label{Eq:W_ADS}
W_{\rm eff}^{\rm ADS}=\left( N_f-N_c\right) \left[ 
\frac{
	{\rm det}\left(T\right) f(Z)
}
{
	\Lambda^{3N_c-N_f}
} 
\right]^{1/(N_f-N_c)}. 
\end{equation}
In this case, $W_{\rm eff}^{\rm ADS}$ vanishes in the classical limit only if $f(Z)$ = 0, 
where the constraint of $BT^{N_f-N_c}{\bar B}$ = ${\rm det}\left(T\right)$, 
namely $Z$ = 1, is satisfied.  The simplest form of $f(Z)$ that satisfies $f(Z)$ = 0 can be given by 
\begin{equation}\label{Eq:f_Z}
f(Z)=( 1-Z )^\rho \ \ (\rho > 0), 
\end{equation}
where $\rho$ is a free parameter. 

If one flavor becomes heavy, our superpotential exhibits a holomorphic decoupling property.  
Add a mass to the $N_f$ flavor, then we have
\begin{equation}\label{Eq:MassiveWeff}
W_{\rm eff}=S 
\left\{ 
\ln\left[
\frac{
	S^{N_c-N_f}{\rm det}\left(T\right) f(Z)
}
{
	\Lambda^{3N_c-N_f}
} 
\right] 
+N_f-N_c\right\} - mT_{N_f}^{N_f}.
\end{equation}
Following a usual procedure, we divide $T$ into ${\tilde T}$ with a light flavor $(N_f-1)$ $\times$ 
$(N_f-1)$ submatrix and $T_{N_f}^{N_f}$ and also $B$ and ${\bar B}$ into light flavored ${\tilde B}$ 
and ${\tilde {\bar B}}$ and heavy flavored parts. At SUSY minimum, the off-diagonal elements of $T$ 
and the heavy flavored $B$ and ${\bar B}$ vanish and $T_{N_f}^{N_f}$ = $S/m$ is derived.  
This relation is referred to as Konishi anomaly relation\cite{Konishi}.  
Inserting this relation into Eq.(\ref{Eq:MassiveWeff}),  we obtain
\begin{equation}\label{Eq:MassiveDecoupledWeff}
W_{\rm eff}=S 
\left\{ 
\ln\left[
\frac{
	S^{N_c-N_f+1}{\rm det}({\tilde T}) f({\tilde Z})
}
{
	{\tilde \Lambda}^{3N_c-N_f+1}
} 
\right] 
+N_f-N_c-1\right\},
\end{equation}
where ${\tilde Z}$ = ${\tilde B}{\tilde T}^{N_f-N_c-1}{\tilde {\bar B}}$/${\rm det}({\tilde T})$ 
from $Z$ = ${\tilde B}T_{N_f}^{N_f}{\tilde T}^{N_f-N_c-1}{\tilde {\bar B}}/T_{N_f}^{N_f}{\rm det}({\tilde T})$ 
and ${\tilde \Lambda}^{3N_c-N_f+1}$ = $m\Lambda^{3N_c-N_f}$.  Thus, after the heavy flavor 
is decoupled at low energies, we are left with Eq.(\ref{Eq:Weff}) with $N_f-1$ flavors.  

We can also derive an effective superpotential for $N_f$ = $N_c-1$ by letting one flavor to be heavy in 
Eq.(\ref{Eq:Weff}) for $N_f$ = $N_c$. For $N_f$ = $N_c$,  at SUSY vacuum, we find that
\begin{equation}\label{Eq:NfNc}
{\rm det}\left(T\right) f(Z) = \Lambda^{2N_c},
\end{equation}
which turns out to be the usual quantum constraint\cite{Seiberg,UsualNfNcConstraint} of
${\rm det}\left(T\right)-B{\bar B}$ = $\Lambda^{2N_c}$ if  $\rho$ = 1 giving $f(Z)$ = $1-Z$. 
The discussion 
goes through in the similar manner to the previous one.  In this case, we find $B$ = ${\bar B}$ = 0, 
leading to $Z$ = 0, and $t$ = $S/m$ at SUSY minimum.  As a result,  Eq.(\ref{Eq:Weff}) 
with $N_f$ = $N_c-1$ is derived if $\Lambda^{2N_c+1}$ is identified with $m\Lambda^{2N_c}/f(0)$, where 
$f(0)$ = 1 by Eq.(\ref{Eq:f_Z}).  The induced $W_{\rm eff}$ is nothing but the famous ADS 
superpotential\cite{EearlySeiberg} after $S$ is integrated out.  It is thus proved that our 
superpotential with Eq.(\ref{Eq:f_Z}) exhibits the holomorphic decoupling property and provides the correct 
superpotential for $N_c$ $<$ $N_f$.  
 
Next, we proceed to discussing what physics is expected in SQCD especially with $N_c+2$ $\leq$ $N_f$ 
$\leq$ $3N_c/2$.  It is known that keeping chiral symmetries unbroken requires the duality description 
using $\ddash$magnetic" quarks.  Therefore, another dynamics, if it is allowed, necessarily 
induces spontaneous breakdown of chiral symmetries.   In our superpotential, Eq.(\ref{Eq:Weff}), 
this dynamical feature is more visible when soft SUSY breaking effects are token into account.  
Although the elimination of $S$ from $W_{\rm eff}$ gives no effects on the SUSY vacuum, 
to evaluate soft SUSY breaking contributions can be well handled by 
$W_{\rm eff}$ with $S$.  Since physics very near the SUSY-invariant vacua is our main concern, all 
breaking masses are kept much smaller than $\Lambda$.  The property of SQCD  
is then inferred from the one examined in the corresponding SUSY-broken vacuum, 
which is smoothly connected to the SUSY-preserving vacuum. 

Let us briefly review what was discussed in Ref.\cite{Yasue} in a slightly different manner. 
To see solely the SUSY breaking effect, we adopt the simplest term that is invariant under 
the chiral symmetries, which is given by the following mass term, ${\cal L}_{mass}$, for the scalar 
quarks, $\phi_A^i$, and antiquarks, ${\bar \phi}^A_i$:
\begin{equation}\label{Eq:SUSYBreakingTerm}
-{\cal L}_{mass}=\sum_{i,A}\left( \mu_L^2\vert \phi_A^i \vert^2 +\mu_R^2\vert {\bar {\phi}}_i^A \vert^2 \right). 
\end{equation}
 Together with the potential term arising from $W_{\rm eff}$, we find that
\begin{eqnarray}
\label{Eq:Veff}
& & V_{\rm eff}  =  G_T \bigg( 
                     \sum_{i=1}^{N_f}  |W_{{\rm eff};i}|^2         
                \bigg)
          + G_B\bigg(
		\sum_{i=B,{\bar B}} |W_{{\rm eff};i}|^2 
       	  \bigg)
         +G_S |W_{{\rm eff};\lambda} |^2 +V_{\rm soft}, \\
\label{Eq:Vsoft}
& & V_{\rm soft}  =  (\mu_L^2 + \mu_R^2 )\Lambda^{-2}\sum_{i=1}^{N_f}  |\pi_i|^2 
                                   +\Lambda^{-2(N_c-1)}
                                   \left( 
                                         \mu_L^2|\pi_B|^2 + \mu_R^2|\pi_{\bar B}|^2
                                   \right)
\end{eqnarray}
with the definition of $W_{{\rm eff};i}$ $\equiv$  $\partial W_{\rm eff}/\partial \pi_i$, etc., 
where $\pi_{\lambda, i,B, {\bar B}}$, respectively, represent the scalar components of 
$S$, $T_i^i$, $B^{[12\cdots N_c]}$ and ${\bar B}_{[12\cdots N_c]}$.  The coefficient 
$G_T$ comes from the K$\ddot{\rm a}$hlar potential, $K$, which is assumed to be diagonal, 
$\partial^2 K/\partial T_i^{k\ast}\partial T_j^\ell$ = $\delta_{ij}\delta_{k\ell}G_T^{-1}$ with 
 $G_T$ = $G_T(T^{\dagger}T)$, and similarly for $G_B$ = $G_B(B^{\dagger}B+{\bar B}^\dagger{
\bar B})$ and $G_S$ = $G_S(S^{\dagger}S)$.  

Since the dynamics requires that some of the $\pi$ acquire non-vanishing VEV's, 
suppose that one of the $\pi_i$ ($i$=1 $\sim$ $N_f$) develops a VEV, and let this be  
labeled by $i$ = $1$: $|\pi_1|$ = $\Lambda_T^2$ $\sim$ $\Lambda^2$.  This VEV  is determined by 
solving  $\partial V_{\rm eff}/\partial\pi_i$ = 0, yielding
\begin{equation}\label{Eq:Tab}
G_TW_{{\rm eff};a}^\ast
\frac{\pi_\lambda}{\pi_a} \left( 1-\alpha_B \right) 
 =  G_SW_{{\rm eff};\lambda}^\ast \left( 1-\alpha_B \right)+\beta_B X_B +M^2\big| \frac{\pi_a}{\Lambda}\big|^2, 
\end{equation}
for $a$=1${\sim}N_c$, where $\alpha_B$ = $zf^\prime (z)/f(z)$ and 
$\beta_B$ = $z\alpha_B^\prime$ with $z$ = $\langle 0|Z|0 \rangle$, and
\begin{eqnarray}
\label{Eq:ScalarMass}
& & M^2 = \mu_L^2+\mu_R^2 +G_T^\prime\Lambda^2\sum_{i=1}^{N_f} \big| W_{{\rm eff};i}\big|^2, \\
\label{Eq:X_B}
& & X_B = G_T\sum_{a=1}^{N_c}W_{{\rm eff};a}^\ast\frac{\pi_\lambda}{\pi_a} -
G_B\sum_{x=B,{\bar B}}W_{{\rm eff};x}^\ast\frac{\pi_\lambda}{\pi_x}. 
\end{eqnarray}
The SUSY breaking effect is specified by $(\mu_L^2+\mu_R^2)\vert \pi_1 \vert^2$ in Eq.(\ref{Eq:Tab}) through $M^2$   
because of $\pi_1$ $\neq$ 0.  This effect is also contained in $W_{{\rm eff};\lambda}$ and 
$X_B$.  From Eq.(\ref{Eq:Tab}), we find that  
\begin{equation}
\bigg| \frac{\pi_a}{\pi_1} \bigg|^2 = 1 + 
\frac{(M^2/\Lambda^2)(\big| \pi_1\big|^2-\big| \pi_a\big|^2)} {G_SW_{{\rm eff};\lambda}^\ast \left( 1-\alpha_B \right) +
(M^2/\Lambda^2)\big| \pi_a\big|^2+\beta_B X_B},
\end{equation}
which cannot be satisfied by $\pi_{a\neq 1}$ = 0.  In fact, $\pi_{a\neq 1}$ = $\pi_1$ is a solution to 
this problem, leading to $|\pi_a|$ = $|\pi_1|$ (= $\Lambda_T^2$). 

Since the classical constraint of $f(z)$ = 0 is expected not to be modified at the SUSY minimum, 
the SUSY breaking effect may arise as tiny deviation of $f(z)$ from 0, which is denoted by 
$\xi$ $\equiv$ $1 - z$ ($\ll$ 1).  By further using the explicit form of Eq.(\ref{Eq:f_Z}) 
for $f(z)$, we find
\begin{equation}
\label{Eq:SUSYbreakingVEV}
\big|\pi_{i=1 \sim N_c} \big|  = \Lambda_T^2, \ \ 
\big|\pi_{i=N_c+1 \sim N_f} \big|  = \xi\big|\pi_{i=1 \sim N_c} \big| ,  \ \
|\pi_B| = |\pi_{\bar B}| \sim \Lambda_T^{N_c}, \ \
\big|\pi_\lambda\big| \sim \Lambda^3\xi^{\frac{\rho+N_f-N_c}{N_f-N_c}},
\end{equation}
in the leading order of $\xi$.  Therefore, in softly broken SQCD, our superpotential 
indicates the breakdown of all chiral symmetries.  This feature is in accord with the 
result of the dynamics of ordinary QCD\cite{QCD}. 

Does the resulting SUSY-breaking vacuum structure persist in the SUSY limit?   
At the SUSY minimum with the suggested vacuum of $|\pi_{a=1\sim N_c}|$ = $\Lambda_T^2$,  
we find the classical constraint of $f(z)$ = 0, as expected, which is derived by 
using $W_{{\rm eff};\lambda}$ = 0 and by noticing that $\pi_\lambda/\pi_{i=N_c+1 \sim N_f}$ = 0 
from $W_{{\rm eff};i}$ = 0. 
In the SUSY limit defined by $\xi$ $\rightarrow$ 0, 
$\pi_{i=N_c+1 \sim N_f}$ vanish to recover chiral $SU(N_f-N_c)$ symmetry and 
$\pi_\lambda$ vanishes to recover chiral $U(1)$ symmetry.  The symmetry breaking is thus  
described by  
\begin{eqnarray}
\label{Eq:ChiralSymmetryBreaking}
& &SU(N_f)_L \times SU(N_f)_R \times U(1)_V \times U(1)_A 
 \nonumber \\
& &\ \ \ \ \ \rightarrow  SU(N_c)_{L+R} \times SU(N_f-N_c)_L \times SU(N_f-N_c)_R \times U(1)^\prime_V \times U(1)^\prime_A, 
\end{eqnarray}
where $U(1)^\prime_V$ is associated with the number
of ($N_f-N_c$)-plet superfields of $SU(N_f-N_c)$ and $U(1)^\prime_A$ is 
associated with the number of $SU(N_c)_{L+R}$ - adjoin and - singlet fermions 
and of scalars of the ($N_f-N_c$)-plet.  
The SUSY vacuum characterized by $|\pi_{a=1\sim N_c}|$ = $\Lambda_T^2$ yields 
spontaneous breakdown of $SU(N_c)_L$ $\times$ $SU(N_c)_R$ to $SU(N_c)_{L+R}$. 
In other words, once the spontaneous breaking is triggered, then $|\pi_{i=1\sim N_c}|$ = 
$\Lambda_T^2$ is a natural solution to SQCD, where soft SUSY breaking can be 
consistently introduced.  
This breaking behavior is translated into the corresponding 
behavior in the Higgs phase by the complementarity\cite{tHooft,Susskind}.  
To generate $SU(N_c)_{L+R}$ can be achieved by $\langle 0| \phi_A^a|0\rangle$ = $\delta_A^a\Lambda_T$ and 
$\langle 0| {\bar \phi}_a^A|0\rangle$ = $\delta_a^A\Lambda_T$, for $a,A$ = 1 $\sim$ $N_c$.  
The anomaly-matching is trivially satisfied in the Higgs phase.  The complementarity shows that massless 
particles are just supplied  by $T_{a=1\sim N_c}^{b=1\sim N_c}$ with Tr($T_a^b$) = 0, 
$T_{i=N_c+1\sim N_f}^{a=1\sim N_c}$, $T^{i=N_c+1\sim N_f}_{a=1\sim N_c}$, $B^{[12\cdots N_c]}$ and 
${\bar B}_{[12\cdots N_c]}$, which are all contained in the Nambu-Goldstone superfields.  
Therefore, the anomaly-matching is automatically satisfied as a result of the spontaneous breakdown and is 
a dynamical consequence.

Let us discuss effects of SUSY-invariant mass terms.  If any quarks with flavors of $SU(N_f-N_c)$ 
are massive, we can determine their vacuum structure from the holomorphic decoupling property.  
On the other hand, if all quarks with flavors of $SU(N_f-N_c)$ are massive, we can further 
utilize instanton contributions\cite{MassiveInstanton} to prescribe vacuum structure.  If our superpotential 
provides correct description of SQCD, both results must be consistent with each other. 

The instanton calculation for the gluino and $N_f$ massless quarks and antiquarks concerns the following 
$SU(N_c)$ - invariant amplitude:
\begin{equation}
\label{Eq:Instanton}
(\lambda\lambda)^{N_c}{\rm det}(\psi^i{\bar \psi}_j), 
\end{equation}
where $\psi ({\bar \psi})$ is a spinor component of $Q$ $({\bar Q})$, which can be converted into
\begin{equation}
\label{Eq:MassiveInstanton}
\prod_{a=1}^{N_c}\pi_a = c\Lambda^{2N_c}\prod_{i=Nc+1}^{N_f}\left(m_i/\Lambda\right), 
\end{equation}
where c ($\neq$ 0) is a coefficient to be fixed.  At our SUSY minimum, the condition of 
$\partial W_{\rm eff}/\partial\pi_\lambda$ = 0 together with 
$\partial W_{\rm eff}/\partial\pi_i$ = 0 for $i$ = $N_c+1$ $\sim$ $N_f$ giving 
$\pi_\lambda/\pi_i$ = $m_i$ reads
\begin{equation}
\label{Eq:MassiveFunc}
f(z) = \prod_{a=1}^{N_c}\left(\Lambda^2/\pi_a\right)\prod_{i=N_c+1}^{N_f}\left(m_i/\Lambda\right).
\end{equation}
By combining these two relations, we observe that the mass dependence in $f(z)$ is completely 
cancelled and derive c = 1/$f(z)$ giving $f(z)\neq$0 instead of $f(z)$ = 0 for the massless 
SQCD. Since $f(z)$ = $(1-z)^\rho$, $z$ $\neq$ 1 is required and the classical
 constraint, corresponding to $z$ = 1,  is modified.  These VEV's are the solution to
\begin{equation}
\label{Eq:MassiveConstraint}
{\rm det}({\tilde T})f({\tilde Z}) = {\tilde \Lambda}^{2N_c},
\end{equation}
where ${\tilde \Lambda}^{2N_c}$ = $\Lambda^{3N_c-N_f}\prod_{i=Nc+1}^{N_f}m_i$, which is the 
quantum constraint (\ref{Eq:NfNc}) for $N_f$ = $N_c$ and which is also consistent with  
the successive use of the holomorphic decoupling.  Therefore, our superpotential for 
$N_c+2$ $\leq$ $N_f$ $\leq$ $3N_c/2$ supplemented by the instanton 
contributions is shown to provide a consistent 
vacuum structure for SQCD with $N_f$ = $N_c$ compatible with the holomorphic decoupling.

A comment is in order for the case with $\rho$ = 1.  
The quantum constraint for $N_f$ = $N_c$, Eq.(\ref{Eq:NfNc}), is rewritten as
\begin{equation}
\label{Eq:NfNcQuantum}
{\rm det}(T)^{1-\rho}\left[{\rm det}(T)-B{\bar B}\right]^\rho = \Lambda^{2N_c}.
\end{equation}
If $\rho$ $\neq$ 1, ${\rm det}(T)$ $\neq$ 0 is required and shows the spontaneous breakdown of chiral 
$SU(N_f)$ symmetry.  There is no room for ${\rm det}(T)$ = 0.  While, if $\rho$ = 1 as in the Seiberg's 
choice\cite{Seiberg}, there are two options: one for the spontaneous breakdown of chiral  
$SU(N_f)$ symmetry and the other for that of vector $U(1)$ symmetry of the baryon number.  The latter case corresponds to 
$z$ = $\infty$, which means that $z\prod_{a=1}^{N_c}\pi_a$ cannot be separated into $z$ and 
$\prod_{a=1}^{N_c}\pi_a$, and is possible to be realized by taking 
\begin{equation}
\label{Eq:Rho1Instanton}
\prod_{a=1}^{N_c}\pi_a = 0, \ \ 
\pi_B\pi_{\bar B} = -\Lambda^{2N_c}\prod_{i=Nc+1}^{N_f}\left(m_i/\Lambda\right)(=-{\tilde \Lambda}^{2N_c}), 
\end{equation}
as instanton contributions.

In summary, we have demonstrated that dynamical symmetry breaking in the $\ddash$electric" SQCD 
with $N_c+2$ $\leq$ $N_f$ $\leq$ $3N_c/2$ ($N_c$ $>$ 2) can be described by
\begin{equation}
\label{Eq:W_eff_summary}
W_{\rm eff}=S 
\left\{ 
\ln\left[
\frac{
	S^{N_c-N_f}{\rm det}\left(T\right) 
	\left(1- Z\right)^\rho
}
{
	\Lambda^{3N_c-N_f}
} 
\right] 
+N_f-N_c\right\} \ \ (\rho > 0)
\end{equation}
with
\begin{equation}
Z = \frac{BT^{N_f-N_c}\bar{B}}{{\rm det}(T)},
\end{equation}
which turns out to be $W_{\rm eff}^{\rm ADS}$ of the ADS-type:
\begin{equation}
W_{\rm eff}^{\rm ADS}=(N_f-N_c) 
\left[
\frac{
	{\rm det}\left(T\right)\left(1-Z\right)^\rho
}
{
	\Lambda^{3N_c-N_f}
} 
\right] ^{1/(N_f-N_c)}.
\end{equation}
This superpotential exhibits 
\begin{enumerate}
\item holomorphic decoupling property,
\item spontaneously breakdown of chiral $SU(N_c)$ symmetry and restoration of chiral $SU(N_f-N_c)$ 
symmetry described by $SU(N_f)_L\times SU(N_f)_R\times  U(1)_V \times U(1)_A 
\rightarrow SU(N_c)_{L+R} \times SU(N_f-N_c)_L \times SU(N_f-N_c)_R \times U(1)^\prime_V 
\times U(1)^\prime_A$,
\item consistent anomaly-matching property due to the emergence of the Nambu-Goldstone superfields, and 
\item correct vacuum structure for $N_f$ = $N_c$ reproduced by instanton contributions when all 
quarks with flavors of $SU(N_f-N_c)$ become massive. 
\end{enumerate}
The breaking of chiral $SU(N_f)$ symmetry to $SU(N_c)_{L+R}$ includes 
the spontaneous breakdown of the vectorial $SU(N_f)_{L+R}$ symmetry, which has also been 
advocated in Ref.\cite{Instability}.  The dependence of the SUSY-breaking effect on various VEV's 
can be summarized as 
\begin{eqnarray}
& 
\big|\langle 0 |T^i_i|0\rangle\big| = 
\left\{ \begin{array}{l}
	 \Lambda_T^2~~~(i=1 \sim N_c) \\
	\xi\Lambda_T^2~~(i=N_c+1 \sim N_f) \\
\end{array} \right.,
\nonumber \\
& \big|\langle 0 |B^{[12\cdots N_c]}|0\rangle\big| = \big|\langle 0 |{\bar B}_{[12\cdots N_c]}|0\rangle\big|  \sim \Lambda_T^{N_c}, \ \ \
\big|\langle 0 |S|0\rangle\big| \sim \Lambda_T^3\xi^{\frac{\rho+N_f-N_c}{N_f-N_c}},
\end{eqnarray}
and the classical constraint of $1-Z$=0 is modified into 
\begin{equation}
1-Z = \xi, 
\end{equation}
where $\xi$ $\rightarrow$ 0 gives the SUSY limit.  

The parameter $\rho$ will be fixed if we find $\ddash$real" properties of SQCD beyond those inferred 
from arguments based on the symmetry principle only.   The choice of 
\begin{equation}
\rho = 1
\end{equation}
seems natural since, in this case, $W_{\rm eff}^{\rm ADS}$ with $N_f$ = $N_c+1$ correctly reproduces 
the Seiberg's superpotential.  Furthermore, the superpotential derived in Ref.\cite{S}:
\begin{equation}
W_{\rm eff}=S 
\left(
\ln
{\cal Z} 
+N_f-N_c+\sum_{n=1}^\infty c_n{\cal Z}^{-n}
\right)
\end{equation}
with 
\begin{equation}
{\cal Z} = \frac{
	S^{N_c-N_f}{\rm det}\left(T\right) 
	\left(1- Z\right)
}
{
	\Lambda^{3N_c-N_f}
}
\end{equation}
has the similar structure to Eq.(\ref{Eq:W_eff_summary}).  This form implies that $\rho$ = 1 
although their additional terms may yield different physics from ours.  

It should be stressed that, in addition to the usually 
believed physics of $\ddash$magnetic" SQCD, where chiral $SU(N_f)$ symmetry is restored, our 
suggested physics of spontaneous chiral symmetry breakdown is expected to be realized in 
$\ddash$electric" SQCD at least for $N_c+2$ $\leq$ $N_f$ $\leq$ $3N_c/2$.  Therefore, 
we expect that there are two phases in SQCD: one with unbroken chiral symmetries realized 
in $\ddash$magneic" SQCD and the other with spontaneously broken chiral symmetries 
realized in $\ddash$electric" SQCD.

\end{document}